\documentclass[english,reprint,amsmaths,amssymb,prl]{revtex4-1}
\usepackage[T1]{fontenc}
\usepackage[latin9]{inputenc}
\usepackage{color}
\usepackage{babel}
\usepackage{array}
\usepackage{refstyle}
\usepackage{units}
\usepackage{multirow}
\usepackage{amsbsy}
\usepackage{amstext}
\usepackage{graphicx}
\usepackage[unicode=true,pdfusetitle,
 bookmarks=false,
 breaklinks=false,pdfborder={0 0 0},pdfborderstyle={},backref=false,colorlinks=true]
 {hyperref}
\hypersetup{
 pdfborderstyle={}}

\makeatletter

\AtBeginDocument{\providecommand\figref[1]{\ref{fig:#1}}}
\AtBeginDocument{\providecommand\tabref[1]{\ref{tab:#1}}}
\AtBeginDocument{\providecommand\eqref[1]{\ref{eq:#1}}}
\providecommand{\tabularnewline}{\\}
\newcommand{\lyxdot}{.}

\RS@ifundefined{subsecref}
  {\newref{subsec}{name = \RSsectxt}}
  {}
\RS@ifundefined{thmref}
  {\def\RSthmtxt{theorem~}\newref{thm}{name = \RSthmtxt}}
  {}
\RS@ifundefined{lemref}
  {\def\RSlemtxt{lemma~}\newref{lem}{name = \RSlemtxt}}
  {}

\usepackage{babel}
\renewcommand{\tabref}{\Tabref}
\renewcommand{\figref}{\Figref}
\usepackage[shortcuts]{extdash}

\makeatother

\begin{document}

\title{Magnetic Multilayer Edges in Bernal-Stacked Hexagonal Boron Nitride}

\author{Mehmet Dogan$^{1,2}$ and Marvin L. Cohen$^{1,2}$}

\affiliation{$^{1}$Department of Physics, University of California, Berkeley,
California 94720, USA $\linebreak$ $^{2}$Materials Sciences Division,
Lawrence Berkeley National Laboratory, Berkeley, California 94720,
USA}
\begin{abstract}
Single-layer \emph{h}-BN is known to have edges with unique magnetism,
however, in the commonly fabricated multilayer $\text{AA}^{\prime}$-\emph{h}-BN,
edge relaxations occur that create interlayer bonds and eliminate
the unpaired electrons at the edge. Recently, a robust method of growing
the unconventional Bernal-stacked \emph{h}-BN (AB-\emph{h}-BN) has
been reported. Here, we use theoretical approaches to investigate
the nitrogen-terminated zigzag edges in AB-\emph{h}-BN that can be
formed in a controlled fashion using a high-energy electron beam.
We find that these ``open'' edges remain intact in bilayer and multilayer
AB-\emph{h}-BN, enabling researchers potentially to investigate these
edge states experimentally. We also investigate the thermodynamics
of the spin configurations at the edge by constructing a lattice model
that is based on parameters extracted from a set of first-principles
calculations. We find that the edge spins in neighboring layers interact
very weakly, resulting in a sequence of independent spin chains in
multilayer samples. By solving this model using Monte Carlo simulations,
we can determine nm-scale correlation lengths at liquid-N$_{2}$ temperatures
and lower. At low temperatures, these edges may be utilized in magnetoresistance
and spintronics applications.
\end{abstract}
\maketitle

\section{Introduction\label{sec:Introduction}}

Among the many novel research avenues that have been created by two-dimensional
(2D) materials, an important one is that their terminations provide
a robust platform to study one-dimensional (1D) systems. For example,
by terminating a sheet of a 2D material by two parallel edges separated
by a small distance, nanoribbons are formed. Similarly, nanoflakes
are made by fabricating three or more edges which fully surround the
nanosheet to obtain a finite (0D) system. These have led to a new
field of research by allowing the study of many physical phenomena
in reduced-dimensional systems \citep{celis2016graphene}. In a similar
vein, fabricating nanopores in 2D materials have been an important
field of research that touches upon molecular sieving, metamaterials
and quantum emission \citep{koenig2012selective,choi2016engineering,zhao2019etching,caldwell2019photonics}.
Nanopores are also made up of edges, but are in a sense the ``opposite''
of nanoflakes. In addition to engineering these edges to fabricate
nanoribbons, nanoflakes and nanopores, edges inevitably occur in real
nanosheets and affect experimental observations. Therefore, it is
important to understand the properties of edges themselves, which
can be done by considering a nanoribbon with a large enough width.
Such a system can be interpreted as a collection of two parallel edges
and an interior region where the edges can be treated independently,
and the interior can be treated like an infinite sheet. In this study,
we investigate the properties of the most commonly observed edge type
in single-layer hexagonal boron nitride (\emph{h}-BN) as well as its
interaction with neighboring layers and edges, which is largely determined
by the stacking sequence.

\begin{figure}
\centering{}\includegraphics[width=0.7\columnwidth]{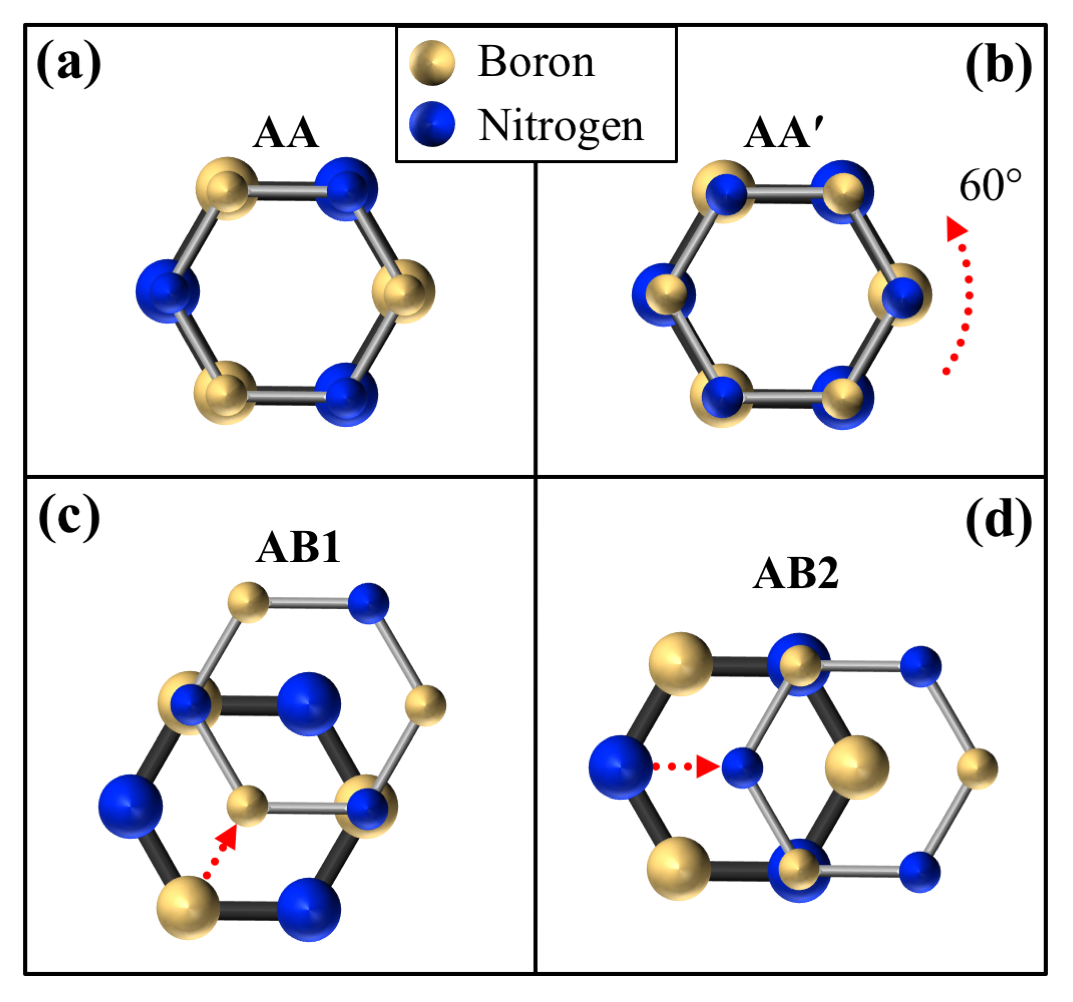}\caption{\label{fig:Stackings}Four high-symmetry stacking sequences of \emph{h}-BN
considered in this study. (a) The AA stacking sequence which has not
been observed experimentally. (b) The commonly observed $\text{AA}^{\prime}$
stacking sequence. (c,d) Two ways of constructing the AB stacking
sequence of \emph{h}-BN. These two ways are physically equivalent
but geometrically distinct, and are distinguished so that the top
and the bottom layers can be treated separately.}
\end{figure}

Stacking the layers of\emph{ h}-BN in different ways yields different
material properties without altering the structure within each layer
\citep{qi2007planarstacking,marom2010stacking,constantinescu2013stacking,gilbert2019alternative}.
The trivial stacking sequence {[}\figref{Stackings}(a){]} is named
the AA stacking, where there is no in-plane shift or rotation between
consecutive layers. When this stacking is repeated to form bulk \emph{h}-BN\emph{,}
columns of B atoms and columns of N atoms in the out-of-plane direction
are created. Although this high-energy stacking sequence has not been
experimentally observed, we include it in our study as a reference,
and find that it has an interlayer distance of $3.64\ \mathring{\text{A}}$.
Common synthesis methods yield a bilayer stacking sequence {[}\figref{Stackings}(b){]},
named the $\text{AA}^{\prime}$ stacking, where there is a $60^{\circ}$
rotation between consecutive layers, which results in columns of alternating
B and N atoms in the bulk \citep{alem2009atomically}. An alternative
stacking sequence, where there is a shift but no rotation between
consecutive layers {[}\figref{Stackings}(c,d){]}, named the AB stacking,
had been observed in rare cases until recently \citep{warner2010atomicresolution,khan2016carbonand,ji2017chemical}.
However, Gilbert \emph{et al.} demonstrated a robust and reproducible
method of growing AB stacked \emph{h}-BN (AB-\emph{h}-BN) \citep{gilbert2019alternative}.
Many computed properties of AB-\emph{h}-BN are similar to those of
$\text{AA}^{\prime}$-\emph{h}-BN, such as interlayer distance (AB:
$3.09\ \mathring{\text{A}}$ vs. $\text{AA}^{\prime}$: $3.11\ \mathring{\text{A}}$),
indirect band gap (AB: 4.43 eV vs. $\text{AA}^{\prime}$: 4.41 eV)
and dielectric tensor \citep{gilbert2019alternative}. Therefore the
new AB-\emph{h}-BN can be used interchangeably with $\text{AA}^{\prime}$-\emph{h}-BN
in many materials applications.

\begin{figure}
\centering{}\includegraphics[width=0.95\columnwidth]{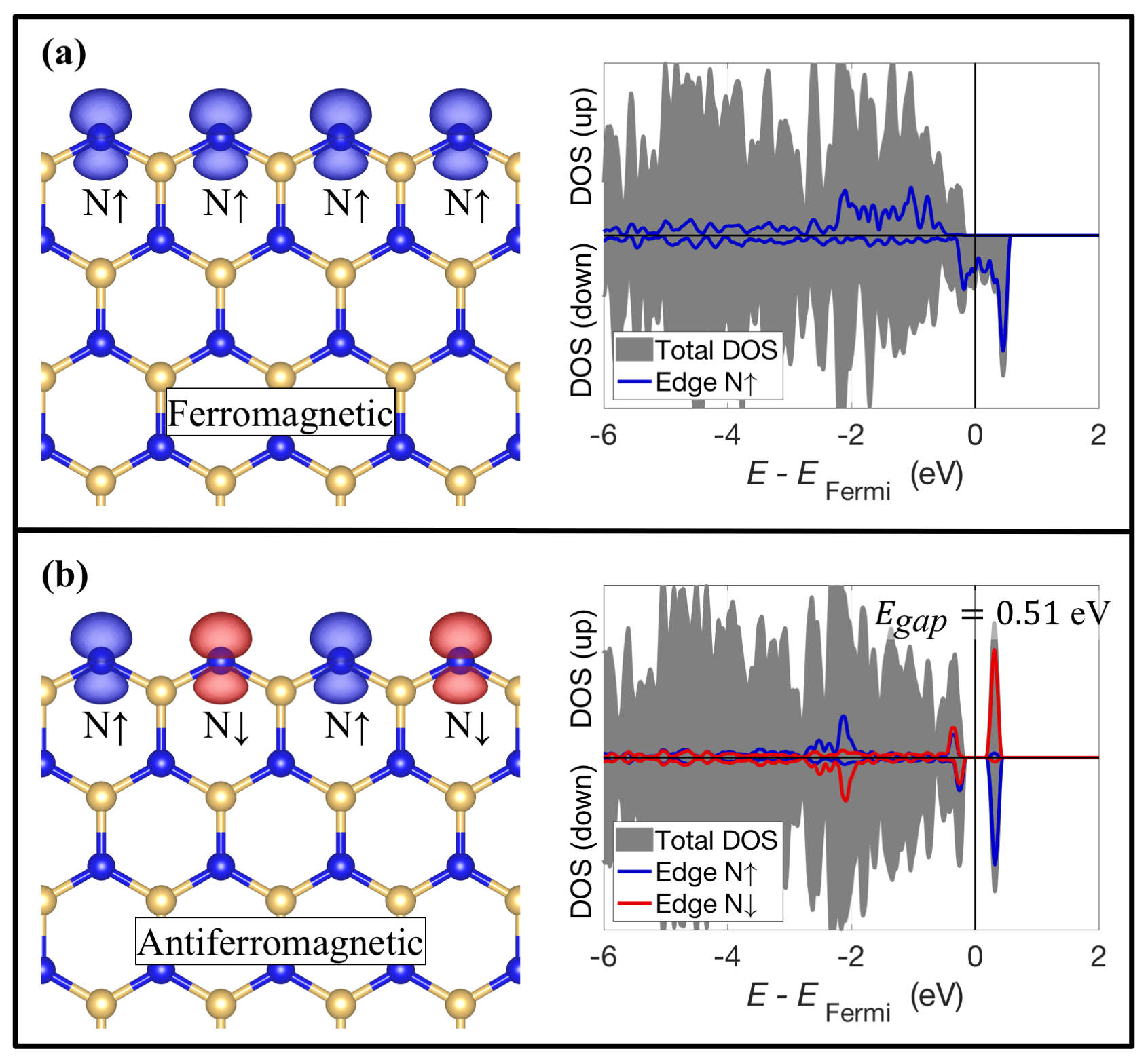}\caption{\label{fig:Monolayer}The ferromagnetic (a) and antiferromagnetic
(b) N-terminated zigzag edge configurations of the monolayer \emph{h}-BN.
The isosurface plots for magnetization with the isovalue $n_{\uparrow}-n_{\downarrow}=\pm0.02\ \left|e\right|/a_{0}^{3}$
are included in the atomic structure pictures. The spin-resolved densities-of-states
plots including the projections onto the orbitals of the edge nitrogen
atom are also shown.}
\end{figure}

Although an infinite sheet of single-layer \emph{h}-BN is a wide-gap
insulator, computational studies have shown that its edges exhibit
a rich collection of electronic and magnetic properties \citep{barone2008magnetic,lai2009magnetic,mukherjee2011edgestabilities,yamijala2013electronic,deng2017theedge}.
Magnetism that is found on these edges arises from unpaired electrons
that occupy dangling $sp^{2}$ hybrid orbitals. However, because of
the existence of various spin configurations within a small total
energy window, a thermodynamic analysis is needed to make predictions
at nonzero temperatures. Furthermore, experimental studies aimed at
locally probing the electronic structure at the edge are needed to
complete our understanding of these systems. The most commonly observed
edge type in \emph{h}-BN is the nitrogen-terminated zigzag edge (N-edge)
\citep{zobelli2007electron,alem2009atomically,meyer2009selective,kotakoski2010electron,kim2011controlled,ryu2015atomicscale,rajan2019addressing,mouhoub2020quantitative},
shown in \figref{Monolayer}. By the application of a high-energy
electron beam in a transmission electron microscopy (TEM) chamber,
it is possible to fabricate the N-edge almost exclusively among all
edge types \citep{alem2011vacancy,pham2016formation,gilbert2017fabrication,gilbert2019alternative,dogan2020electron}.
In this study, we investigate the spin configurations of the N-edge
in single-layer \emph{h}-BN and build a statistical model to make
predictions about the magnetic properties of real edges. We then discuss
what happens to the N-edge in multilayer \emph{h}-BN. Because of the
lack of relative rotation between the consecutive layers in AB-\emph{h}-BN,
multilayer N-edges can be formed \citep{gilbert2019alternative,dogan2020electron},
which is not possible in $\text{AA}^{\prime}$-\emph{h}-BN because
of the $60^{\circ}$ rotation between consecutive layers \citep{alem2011vacancy}.
Additionally, it has been observed that in $\text{AA}^{\prime}$-\emph{h}-BN,
edge relaxations occur and give rise to covalent bonding between the
edge atoms in the neighboring layers \citep{alem2012subangstrom},
which is not the case for AB-\emph{h}-BN \citep{dogan2020electron}.

We note that in AB-\emph{h}-BN, the two consecutive layers are inequivalent.
Specifically, in \figref{Stackings}(c), the N atom in the top layer
is aligned with the B atom in the bottom layer, whereas the B atom
in the top layer is aligned with the hollow site of the bottom layer.
Therefore the B (N) atom of the top layer is not in the same physical
environment as the B (N) atom of the bottom layer. Given a starting
bottom layer, the top layer may be formed in two distinct ways, shown
in \figref{Stackings}(c,d), which we name AB1 and AB2, respectively.
The top layer in AB1 then becomes equivalent to the bottom layer in
AB2, and vice versa. So the distinction between these two types within
the AB stacking is meaningful when we are considering a well-defined
bottom layer and a well-defined top layer, as we will in the discussion
below.

\section{Methods\label{sec:Methods}}

We use density functional theory (DFT) in the Perdew\textendash Burke\textendash Ernzerhof
generalized gradient approximation (PBE GGA) to conduct our \emph{ab
initio} calculations \citep{perdew1981selfinteraction}. We employ
the QUANTUM ESPRESSO software package with norm-conserving pseudopotentials
\citep{giannozzi2009quantum,hamann2013optimized}. We have converged
our calculations with respect to the plane-wave energy cutoff for
the pseudo Kohn-Sham wavefunctions, which has yielded a value of 80
Ry. For single- and double-layer edge simulations, we use a $12\times1\times1$
Monkhorst\textendash Pack k-point mesh per unit edge to sample the
Brillouin zone \citep{cohen1975selfconsistent}. A $1\times10$ cell
is constructed, and $\sim$12 Å of vacuum is placed between the copies
of the 1D system along the direction perpendicular to the edge. The
dangling bonds at the opposite edge are passivated using hydrogen
atoms. We keep the first 4 unit cells (8 atoms) unrelaxed, and relax
the remaining 6 unit cells (12 atoms). In order to include the interlayer
van der Waals interactions, we include a Grimme-type dispersion correction
\citep{grimme2006semiempirical}. All atomic coordinates are relaxed
until the forces on all the atoms are less than $10^{-3}$ Ry/$a_{0}$
in all three Cartesian directions, where $a_{0}$ is the Bohr radius.
A $\sim$14 Å thick vacuum is used between the periodic copies of
the slab in the out-of-plane direction to isolate the sheets.

\section{Results\label{sec:Results}}

\subsection{Open Edges in \emph{h}-BN\label{subsec:Edges1}}

In \figref{Monolayer}, we present the N-edge of a monolayer \emph{h-}BN
sheet. The ferromagnetic (FM) and antiferromagnetic (AFM) states are
presented in \figref{Monolayer}(a,b), respectively. The isosurface
plots for magnetization with the isovalue $n_{\uparrow}-n_{\downarrow}=\pm0.02\ \left|e\right|/a_{0}^{3}$
show a spatially localized spin polarization on the dangling $sp^{2}$
orbitals on the edge N atoms. Both states are lower in energy than
the nonmagnetic state, and the ferromagnetic state is the ground state
{[}\tabref{Edges}{]}. In both the FM and AFM states, the magnetization
per edge N atom is $1$ $\mu_{\text{B}}$. There is negligible relaxation
($<0.01\ \mathring{\text{A}}$) at these edges arising from the change
in the spin configuration. The spin-resolved densities-of-states (DOS)
plots show that the ferromagnetic edge is a half-metal, i.e. the majority
spin channel is insulating and the minority spin channel is a metal,
which should give rise to a fully spin-polarized edge current and
potentially enable spintronic devices such as tunnel junctions, spin
filters, diodes and transistors \textcolor{black}{\citep{prinz1998magnetoelectronics,wolf2001spintronics,vzutic2004spintronics,felser2007spintronics}}.
The antiferromagnetic state is a semiconductor with a gap of 0.51
eV. Because the spin configuration and the electronic structure are
closely linked, these edges may be useful in metal spintronics as
well as semiconductor spintronics applications\textcolor{black}{{} \citep{prinz1998magnetoelectronics,wolf2001spintronics,vzutic2004spintronics,awschalom2007challenges,felser2007spintronics,yazyev2010emergence,awschalom2013quantum}.}
Our findings on the monolayer \emph{h}-BN agree well with the existing
computational studies \citep{barone2008magnetic,du2009dotsversus,lai2009magnetic,si2014intrinsic,deng2017theedge}.

When the bilayer \emph{h}-BN is formed in the AB stacking, the N-edge
remains ``open'' upon relaxation, \emph{i.e.} no interlayer bonds
form. This results in essentially two copies of the monolayer \emph{h}-BN
sheet including the edge. This bilayer configuration can be repeated
along the out-of-plane direction to form a bulk N-edge. The total
energies of the FM state and the AFM state (denoted AFM$\uparrow\downarrow$)
for the bilayer and bulk N-edge are listed in \tabref{Edges}. The
FM state remains the ground state and is preferred to the AFM$\uparrow\downarrow$
state by a similar difference in energy, compared to that of the monolayer
\emph{h}-BN. If the consecutive layers are each ferromagnetic but
spin-polarized in the opposite direction with respect to each other
(denoted AFM$_{\downarrow}^{\uparrow}$), the total energy is slightly
higher than the FM case. All of these results reinforce our finding
that the neighboring layers are electronically and magnetically largely
independent from each other.

\begin{table}
\def\arraystretch{2.0}
\begin{centering}
\begin{tabular}{c|c|c|c|c|c|}
\multicolumn{2}{c|}{} & NM & FM & AFM$\uparrow\downarrow$ & AFM$_{\downarrow}^{\uparrow}$\tabularnewline
\hline 
\multirow{2}{*}{Monolayer} & $\Delta E$ (meV) & $\equiv0$ & \multicolumn{1}{c|}{$-218$} & $-184$ & \multirow{2}{*}{N/A}\tabularnewline
\cline{2-5} 
 & $E_{gap}$ (eV) & metal & \multicolumn{1}{c|}{half-metal} & $0.51$ & \tabularnewline
\hline 
\multirow{2}{*}{Bilayer (AB)} & $\Delta E$ & $\equiv0$ & $-224$ & $-187$ & $-222$\tabularnewline
\cline{2-6} 
 & $E_{gap}$ & metal & half-metal & $0.42$ & $0.18$\tabularnewline
\hline 
\multirow{2}{*}{Bulk (AB)} & $\Delta E$ & $\equiv0$ & \multicolumn{1}{c|}{$-226$} & $-195$ & \multicolumn{1}{c|}{$-207$}\tabularnewline
\cline{2-6} 
 & $E_{gap}$ & metal & \multicolumn{1}{c|}{half-metal} & $0.31$ & \multicolumn{1}{c|}{metal}\tabularnewline
\hline 
\end{tabular}
\par\end{centering}
\caption{\label{tab:Edges}Total energies and electronic band gaps of the nonmagnetic
(NM), ferromagnetic (FM) and antiferromagnetic (AFM) configurations
of the N-terminated zigzag edge of \emph{h}-BN. Total energies are
given per edge N atom.}
\end{table}

\begin{figure*}
\centering{}\includegraphics[width=0.85\textwidth]{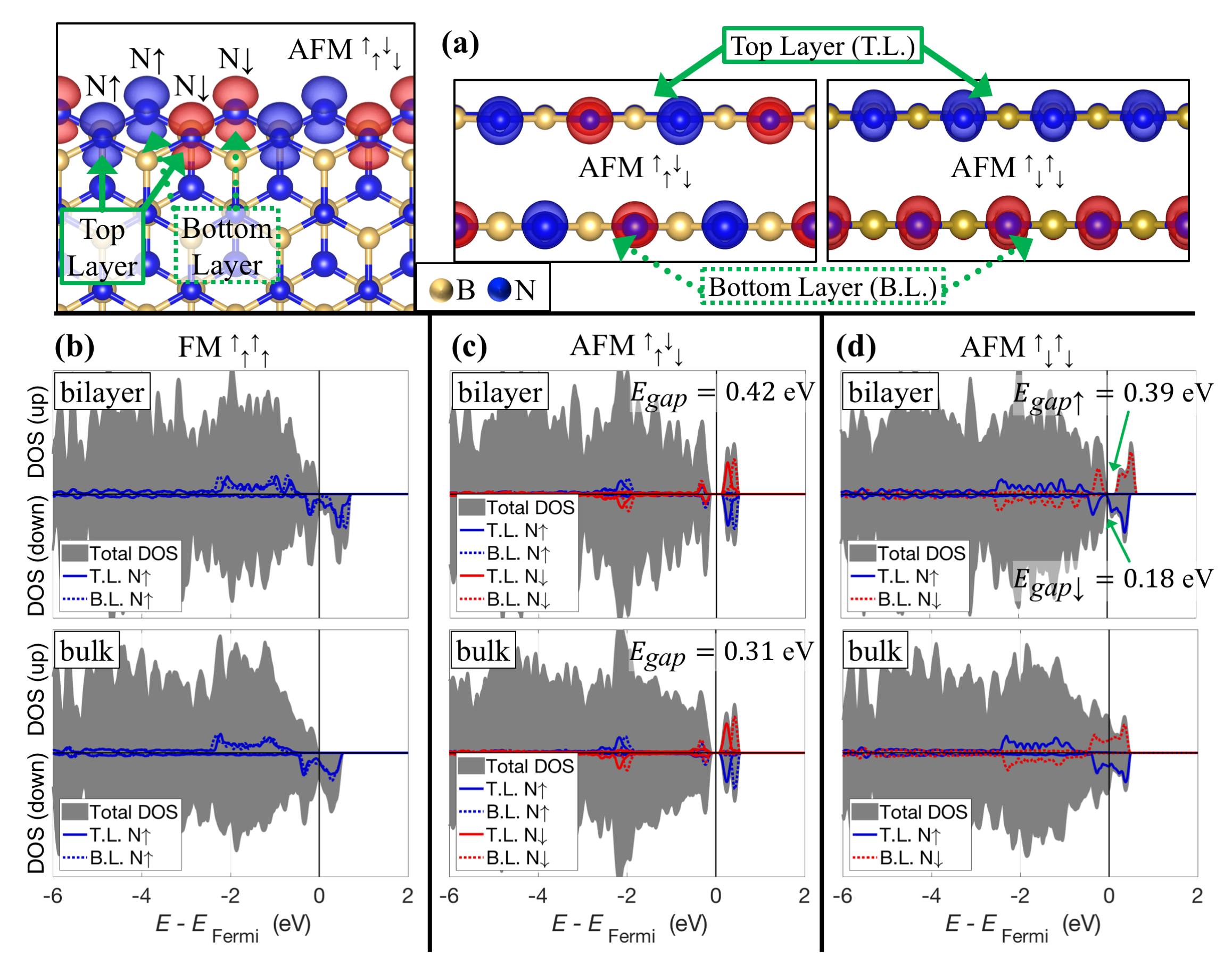}\caption{\label{fig:Bilayer_mag}Atomic and electronic structure of the bilayer
N-edge in AB-\emph{h}-BN. (a) The isosurface plots for magnetization
with the isovalue $n_{\uparrow}-n_{\downarrow}=\pm0.02\ \left|e\right|/a_{0}^{3}$
are included in the atomic structure pictures for the two antiferromagnetic
(AFM) configurations. (b) The spin-resolved densities-of-states plots
including the projections onto the orbitals of the edge nitrogen atoms
for the ferromagnetic (FM) configuration. (c, d) Same as (b), but
for the two AFM configurations. The values of the electronic gaps,
where they exist, are also printed on each panel.}
\end{figure*}

In \figref{Bilayer_mag}, the details of the atomic and magnetic configuration
of the bilayer and bulk AB1-\emph{h}-BN N-edge are presented. In \figref{Bilayer_mag}(a),
the isosurface plots for magnetization (isovalue $n_{\uparrow}-n_{\downarrow}=\pm0.02\ \left|e\right|/a_{0}^{3}$
) for the two AFM configurations of the bilayer edge are displayed
(AFM$_{\ \uparrow\ \downarrow}^{\uparrow\ \downarrow}$=AFM$\uparrow\downarrow$
and AFM$_{\ \downarrow\ \downarrow}^{\uparrow\ \uparrow}$=AFM$_{\downarrow}^{\uparrow}$).
In \figref{Bilayer_mag}(b,c,d), the DOS plots show that the bilayer
FM edge is a half-metal, the bilayer AFM$\uparrow\downarrow$ state
is a semiconductor, the bilayer AFM$_{\downarrow}^{\uparrow}$ state
is a half-semiconductor, and the bulk AFM$_{\downarrow}^{\uparrow}$
state is a metal. The bilayer AFM$_{\downarrow}^{\uparrow}$ state
is similar to a bipolar magnetic semiconductor in which the valence
band maximum (VBM) and the conduction band minimum (CBM) of one spin
channel are at a higher energy than the those of the other spin channel
\citep{li2012bipolar,li2016firstprinciples}. Here, the VBMs of the
two spin channels are aligned but the CBMs are not aligned, leading
to unequal gaps in the two spin channels, which, in the case of slight
electron doping, would give rise to a spin-polarized edge current
that would also be spatially confined to the top layer. Manipulating
spins in semiconductors (semiconductor spintronics) is a rapidly developing
field with promising applications in quantum computing \citep{awschalom2007challenges,awschalom2013quantum}.

Because the energy differences between the FM and AFM configurations
are low, thermodynamic considerations are needed to determine the
expected configurations at a given temperature. Because spin polarization
is localized to each edge N atom, we can use a discrete lattice model
with interactions between neighbors instead of an itinerant spin model.
In order to extract the interaction strengths, we have computed the
energies of 10 single-layer, 5 bilayer and 5 bulk spin configurations.
Because of the high precision required from these calculations, only
the supercells compatible with the $12\times1$ $k$-point grid are
considered, which are $1\times1$, $2\times1$, $3\times1$, $4\times1$
and $6\times1$. The spin configurations, total energies and the electronic
gaps of these systems are listed in Table S1 of the Supplemental Material.
These calculations have allowed us to extract the in-plane interaction
energies for nearest, next-nearest and next-next-nearest neighbors
($J_{1}$, $J_{2}$ and $J_{3}$, respectively), as well as the out-of-plane
interaction energy ($J_{4}$). The schematic representations of these
interactions are presented in Figure S1, and the fitted values with
the 90\% confidence intervals are listed in Table S2 of the Supplemental
Material. We find that zero is within the confidence interval for
$J_{3}$ and $J_{4}$, meaning that the open N-edge of each layer
can be thought of as an independent one-dimensional spin chain with
nearest and next-nearest neighbor interactions with the Hamiltonian:\textcolor{black}{
\begin{equation}
H=J_{1}\sum_{i}\boldsymbol{S_{i}}\cdot\boldsymbol{S_{i+1}}+J_{2}\sum_{i}\boldsymbol{S_{i}}\cdot\boldsymbol{S_{i+2}}-\mu_{\text{B}}H\sum_{i}S_{i}^{z},\label{eq:Hamiltonian}
\end{equation}
where $\boldsymbol{S_{i}}$ is a unit vector with $x,y,z$ components
that represents the spin at a lattice site $i$, $J_{1}=-17$ meV
is the nearest neighbor interaction energy, $J_{2}=-12$ meV is the
next-nearest neighbor interaction energy, $\mu_{\text{B}}$ is the
Bohr magneton, and $H$ is the external magnetic field. Because spin\textendash orbit
interaction in }\textcolor{black}{\emph{h}}\textcolor{black}{-BN is
extremely weak \citep{zollner2019heterostructures} (two or three
orders of magnitude smaller than $J_{1}$ and $J_{2}$), we do not
expect magnetic anisotropy in this system above $\sim10$ K \citep{yazyev2008magnetic}.
As a result of this lack of anisotropy, there is no easy axis present
in this system which leads to a continuous spin model (Heisenberg)
instead of a discrete spin model (Ising); furthermore, the external
magnetic field can be assumed to apply in the $z$-direction without
loss of generality.}

\textcolor{black}{This model corresponds to classical one-dimensional
isotropic Heisenberg model with nearest and next-nearest neighbor
interactions. For such a model, according to the Mermin\textendash Wagner
theorem, in the absence of an external magnetic field, the ferroelectric
ground state is not stable for any temperature greater than absolute
zero \citep{mermin1966absence}. Although an exact analytical solution
to this model is not known, when both $J_{1}$ and $J_{2}$ are negative,
}\textcolor{black}{\emph{i.e.}}\textcolor{black}{{} both the interactions
are ferromagnetic, its behavior is analogous to the Heisenberg model
with only nearest-neighbor interactions \citep{majumdar1969onnextnearestneighbor,majumdar1969onnextnearestneighbor2}
whose exact solution is known \citep{fisher1964magnetism}. In order
to determine the behavior of the expected value of spin ($\left\langle S_{z}\right\rangle $)
and the spin pair correlation function ($\xi_{z}$) in our system
where $J_{1}$ and $J_{2}$ have specific numerical values, we have
conducted a Monte Carlo study a modified version of the Metropolis
algorithm \citep{metropolis1953equation}. In the usual Metropolis
algorithm for the Heisenberg model, at each simulation step, an attempt
is made to randomize one of the spins \citep{slanic1991dynamics}.
However, this leads to a critical slowing down where the decorrelation
time diverges near the critical temperature \citep{barkema1997newmonte,dogan2019theoryof},
which in this case is 0 K. To overcome this problem to some degree,
instead of choosing the proposed direction of the spin at random,
we choose it according to the formula 
\begin{equation}
\Delta\theta=\cos^{-1}\left(K^{-1}\log(\exp(-K)+2r\sinh(K))\right),
\end{equation}
where $\Delta\theta$ is the angle between the new spin and the old
spin, $K=\nicefrac{\left(J_{1}+J_{2}\right)}{\left(k_{\text{B}}T\right)}$
and $r$ is a random number between 0 and 1 \citep{lurie1974computer}.
This attempt is then accepted or rejected based on the Boltzmann factor
of the energy difference between the proposed and original configurations,
which generates a sampling of the configuration space that obeys the
detailed balance condition \citep{dogan2019theoryof}.}

\begin{figure}
\centering{}\includegraphics[width=0.85\columnwidth]{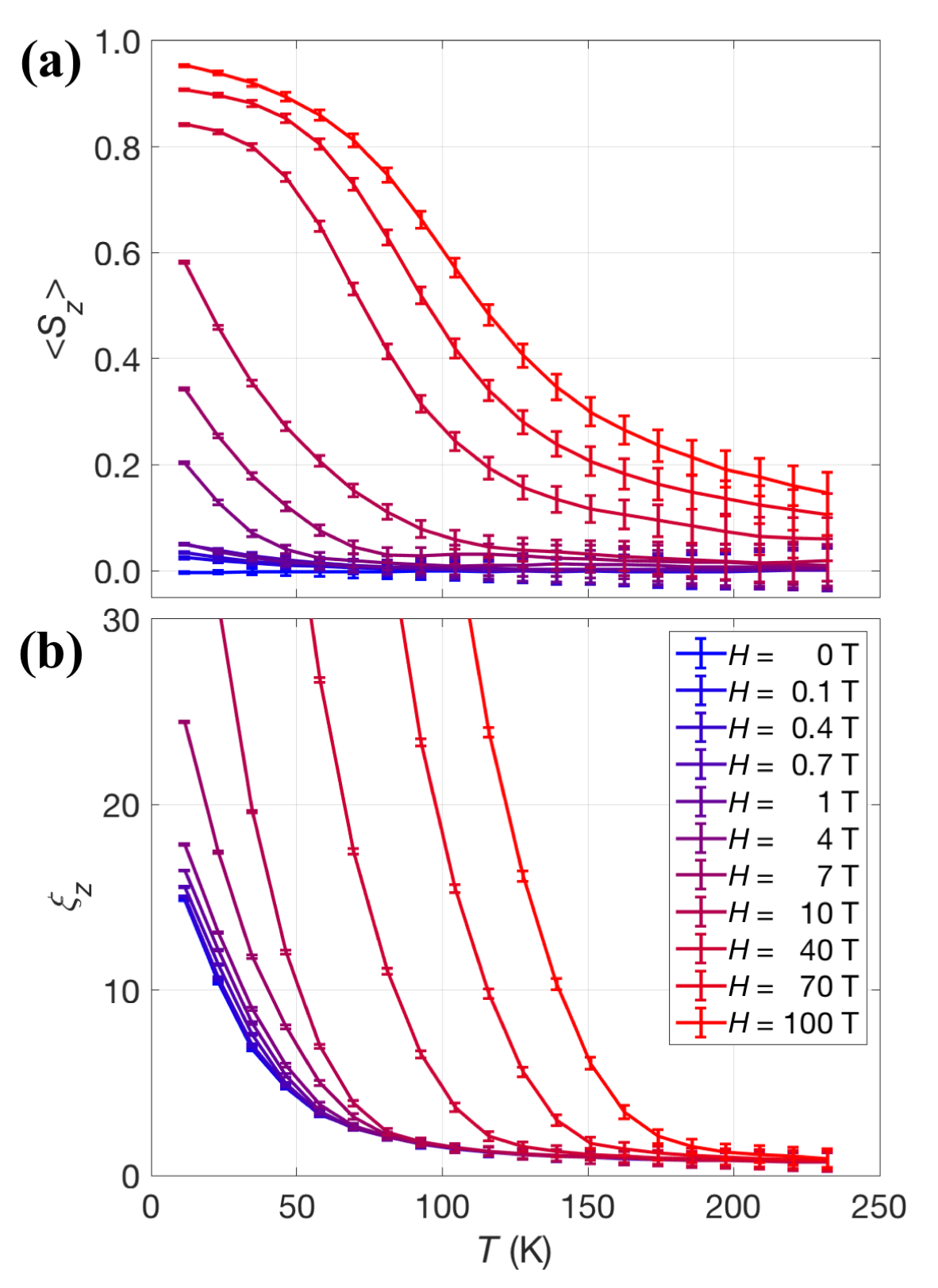}\caption{\label{fig:Heisenberg_res} \textcolor{black}{The expected value of
spin ($\left\langle S_{z}\right\rangle $) (a) and spin pair correlation
function ($\xi_{z}$) (b) for all temperatures and magnetic fields.
The error bars are obtained from the standard deviations over ten
simulations conducted for each $H,T$ pair.}}
\end{figure}

We have obtained a collection of results by running these simulations
with free boundary conditions where the number of spins is chosen
between $N=20$ and $N=200$ to be at least a few times the spin pair
correlation length, the temperature varies between $T=12$ K and $T=232$
K and the external magnetic field varies between $H=0$ T and $H=100$
T. As an example, the correlation functions and the evolution of the
average site energy for a simulation for $H=0.1$ T and $T=70$ K
are presented in Figures S2, S3 and S4 of the Supplemental Material.
\textcolor{black}{$\left\langle S_{z}\right\rangle $ and $\xi_{z}$
for all temperatures and magnetic fields are plotted in \figref{Heisenberg_res}.
The error bars are obtained from the standard deviations over ten
simulations conducted for each $H,T$ pair. According to our results
on $\left\langle S_{z}\right\rangle $ {[}\figref{Heisenberg_res}(a){]},
it is possible to magnetize significantly these edges with external
fields of 4 T and above. This is higher than the field for magnetic
saturation (about 1 T) found in an experiment \citep{si2014intrinsic}.
However, in this experiment, the observed magnetization may have originated
from not only edges but also other defects such as vacancies and pores,
which are also known to have magnetic properties \citep{dogan2020electron}.
Although the average spin for $H=0$ T is zero for all nonzero temperatures,
the spin pair correlation length is not zero {[}\figref{Heisenberg_res}(b){]}
and increases exponentially as $T$ approaches zero. As a result,
the expectation value of the site energy decreases as $T\rightarrow0$
{[}Figure S5 }of the Supplemental Material\textcolor{black}{{]}. At
low-temperature experiments (liquid N$_{2}$ temperatures or lower),
correlation lengths comparable to the edge length may produce an enhanced
magnetization in the presence of small and intermediate-sized nanopores.
The heat capacity ($C_{v}$) and magnetic susceptibility ($\chi$)
of this system for all $H,T$ pairs are also presented in Figures
S6 and S7 of }the Supplemental Materia\textcolor{black}{l, whose behaviors
are analogous to similar models \citep{bonner1978onedimensional}.
Experiments that can isolate the magnetic behavior of edges in }\textcolor{black}{\emph{h}}\textcolor{black}{-BN
sheets, such as electron energy loss spectroscopy (EELS) \citep{alem2012subangstrom}
and scanning tunneling microscopy (STM) \citep{wong2015characterization}
are needed to test these findings. The ability to induce the ferromagnetic
state at the edge }\textcolor{black}{\emph{via}}\textcolor{black}{{}
an external field, which is also the state with less electrical resistance
(\figref{Monolayer}), indicates the potential for magnetoresistive
effect \citep{prinz1998magnetoelectronics,wolf2001spintronics,felser2007spintronics}.
}Furthermore, if edges could be stacked so that FM and AFM edges alternate\textcolor{black}{,
}giant magnetoresistance may be observed \textcolor{black}{\citep{prinz1998magnetoelectronics,wolf2001spintronics,felser2007spintronics}}. 

\subsection{Closed Edges in \emph{h}-BN\label{subsec:Edges2}}

\begin{figure*}
\centering{}\includegraphics[width=0.75\textwidth]{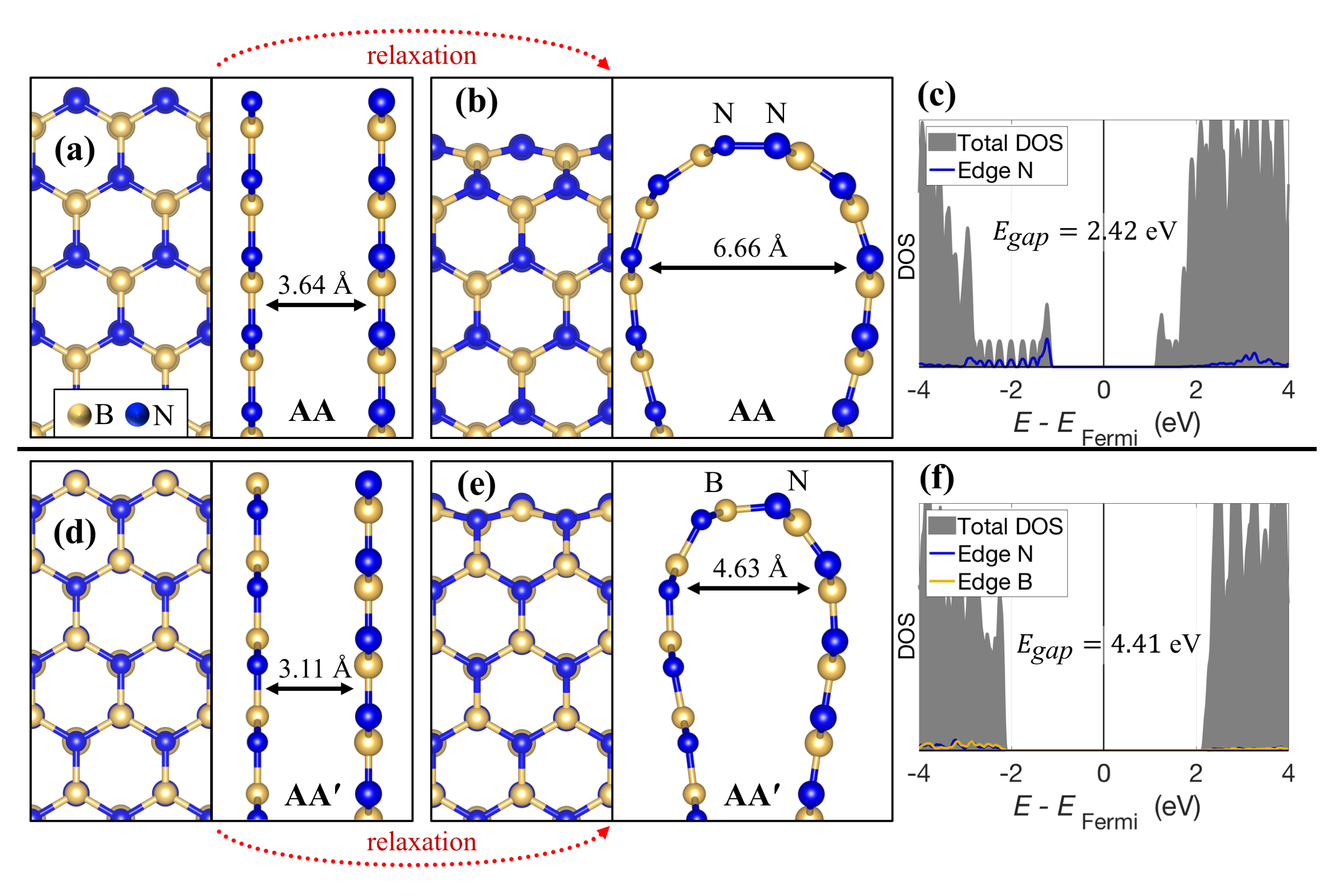}\caption{\label{fig:Bilayer_AA}Edge reconstructions of AA- and $\text{AA}'$-\emph{h}-BN.
Top and side views of the atomic structure before and after the relaxation
is presented for AA-\emph{h}-BN (a,b) and $\text{AA}'$-\emph{h}-BN
(d,e). The densities-of-states plots including the projections onto
the orbitals of the edge atoms marked in the middle panels are presented
for AA-\emph{h}-BN (c) and $\text{AA}'$-\emph{h}-BN (f).}
\end{figure*}

We now turn to other possibilities of bilayer edge reconstructions
in \emph{h}-BN which result in ``closed'' edges. First, as reference,
we compute the relaxation of the N-edge in AA- and $\text{AA}'$-\emph{h}-BN.
In AA-\emph{h}-BN, edge N atoms in neighboring layers are directly
aligned, and relax out-of-plane to form an $sp^{2}$ type bond {[}\figref{Bilayer_AA}(a,b){]}.
The degree of out-of-plane relaxation is quite large, where the highest
interlayer distance becomes $6.66$ Å. In $\text{AA}'$-\emph{h}-BN,
due to the inequivalence between the neighboring layers, an N-edge
in the bottom layer must align with a B-edge in the top layer. These
edges form an interlayer bond which creates a structure that can be
understood as a half BN nanotube {[}\figref{Bilayer_AA}(d,e){]}.
Our findings on the closed $\text{AA}'$-\emph{h}-BN edge agrees with
previous calculations and observations \citep{alem2012subangstrom}.
In both of these edge reconstructions, each atom makes three bonds,
which eliminates unpaired electrons and results in nonmagnetic insulators
{[}\figref{Bilayer_AA}(c,f){]}.

\begin{figure*}
\centering{}\includegraphics[width=0.75\textwidth]{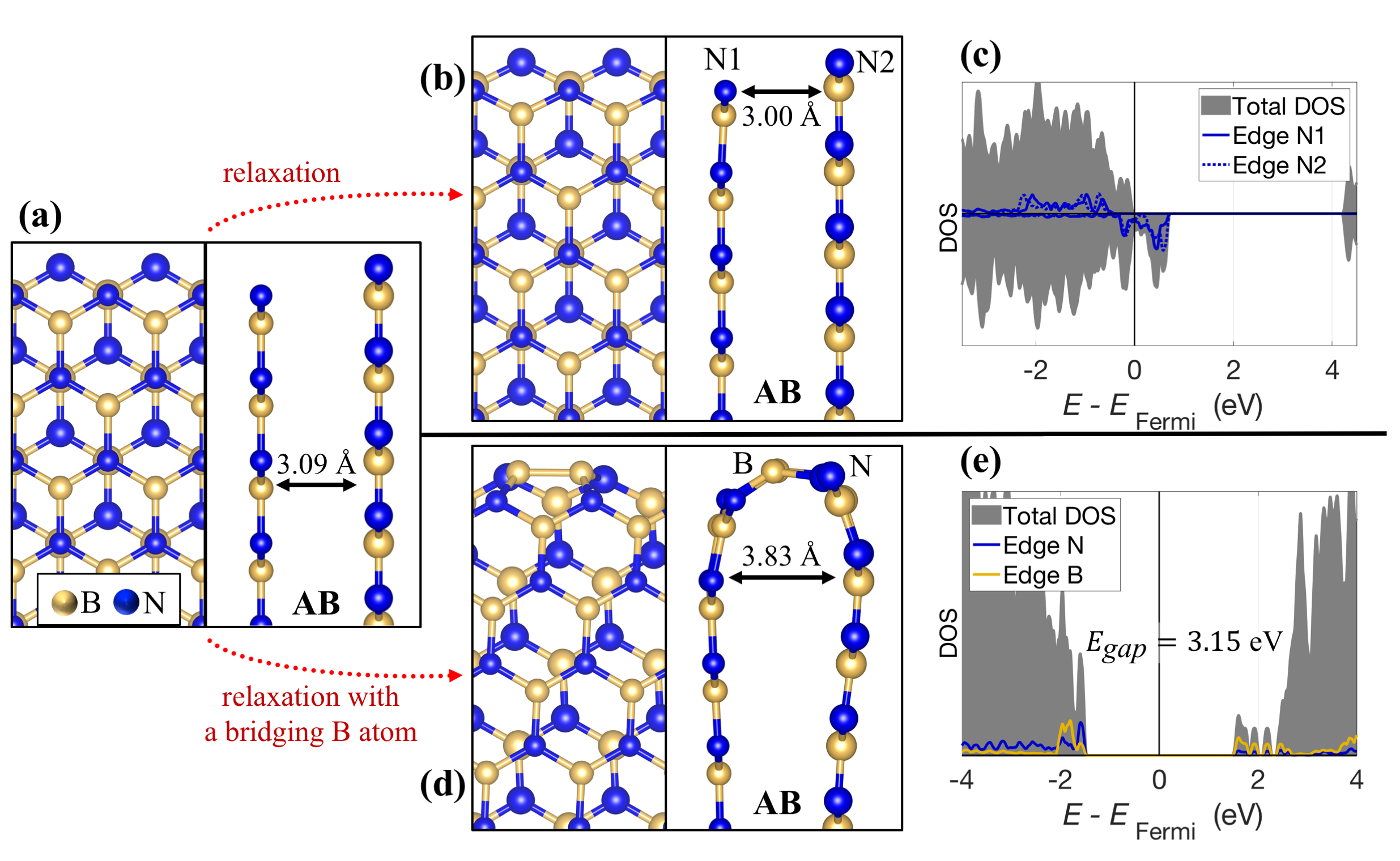}\caption{\label{fig:Bilayer_AB1}Edge reconstructions of AB-\emph{h}-BN arising
from the first starting configuration. Top and side views of the atomic
structure before (a) and after the relaxation (b) are presented, along
with the spin-resolved densities-of-states plot (c) including the
projections onto the orbitals of the edge atoms marked in the middle
panel. The structure that results from adding a bridging B atom between
the two edges is presented in (d) along with its densities-of-states
plot (e) including the projections onto the orbitals of the edge atoms
marked in the middle panel.}
\end{figure*}

In the case of AB-\emph{h}-BN, the starting configuration of the bilayer
N-edge can be created in two different ways by terminating the layers
at different N rows. The first is presented in \figref{Bilayer_AB1}(a).
When the atoms are allowed to relax, no interlayer bonds form, leading
to the open edge structure {[}\figref{Bilayer_AB1}(b,c){]} which
we have discussed above. This bilayer edge is the one we were able
to identify in the HRTEM study \citep{dogan2020electron}. However,
it is conceivable that while the edges form, the B atom bonded to
the edge N atom in the top layer finds the edge N atom in the bottom
layer and forms a bridge between the two edges. This reconstruction
is shown in \figref{Bilayer_AB1}(d). This highly distorted $2\times1$
edge structure has not been observed, but may arise under different
experimental conditions. Because each atom makes three bonds, eliminating
unpaired electrons, the DOS plot shows a nonmagnetic insulator {[}\figref{Bilayer_AB1}(e){]}.

\begin{figure*}
\centering{}\includegraphics[width=0.75\textwidth]{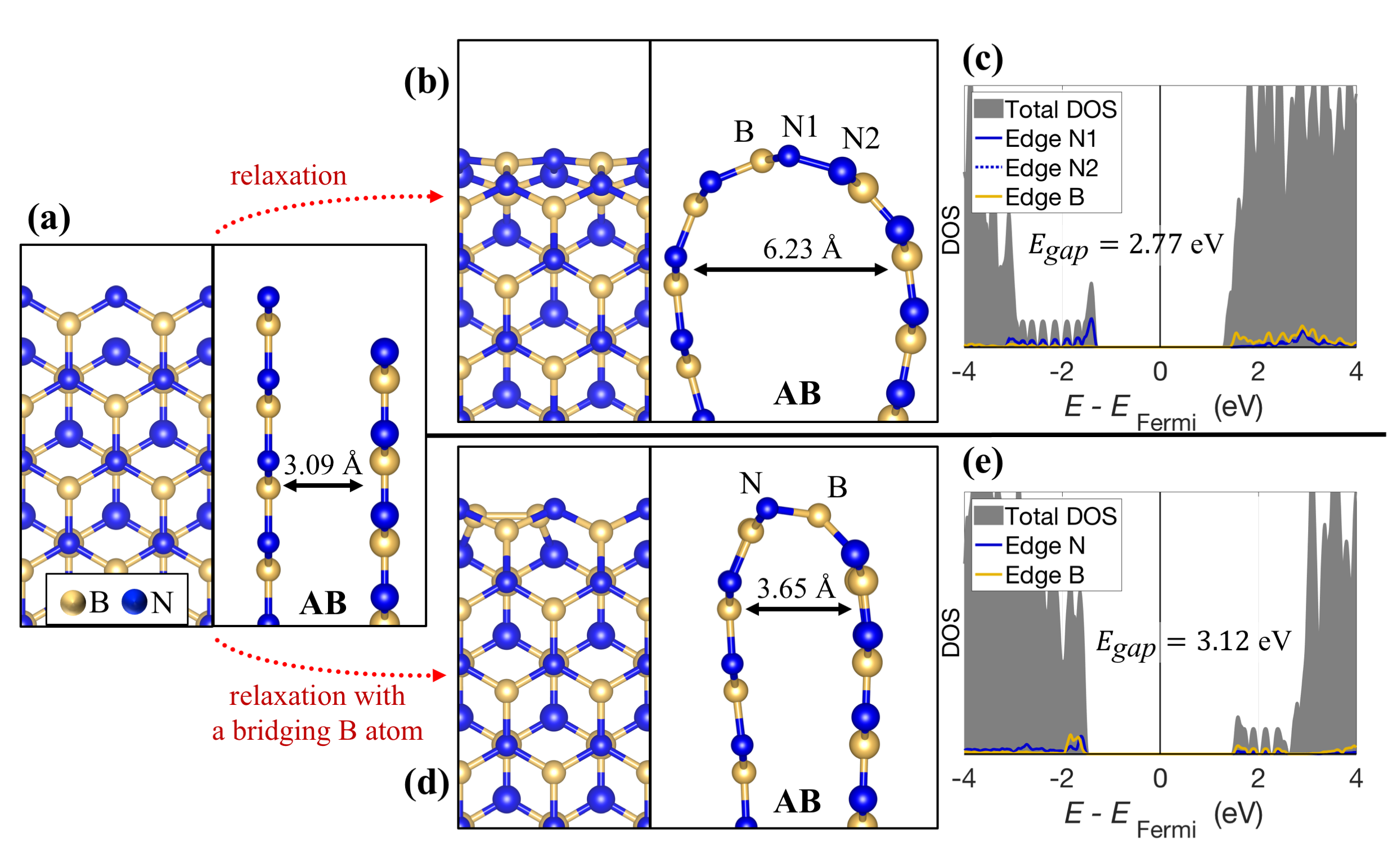}\caption{\label{fig:Bilayer_AB2}Edge reconstructions of AB-\emph{h}-BN arising
from the second starting configuration. Top and side views of the
atomic structure before (a) and after the relaxation (b) are presented,
along with the densities-of-states plot (c) including the projections
onto the orbitals of the edge atoms marked in the middle panel. The
structure that results from adding a bridging B atom between the two
edges is presented in (d) along with its densities-of-states plot
(e) including the projections onto the orbitals of the edge atoms
marked in the middle panel.}
\end{figure*}

The second way of creating a starting configuration for bilayer N-edge
in AB-\emph{h}-BN is presented in \figref{Bilayer_AB2}(a). When the
atoms are allowed to relax, large out-of-plane relaxations occur,
leading to the closed edge structure shown in \figref{Bilayer_AB2}(b).
The large out-of-plane relaxations are accompanied by a significant
in-plane contraction perpendicular to the edge, which was not observed
in the HRTEM study \citep{dogan2020electron}. It is possible that
while the edges form, the B atom bonded to the edge N atom in the
bottom layer finds the edge N atom in the top layer and forms a bridge
between the two edges. This reconstruction is shown in \figref{Bilayer_AB2}(d).
Although this forms a closed edge with a $2\times1$ reconstruction,
the in-plane relaxations are strictly confined to the edge atoms and
no significant contraction perpendicular to the edge is observed.
As a result, this structure would be very difficult to identify in
an HRTEM study such as Ref. \citealp{dogan2020electron}, and therefore
is not ruled out. In both of these reconstructions, because each atom
makes three bonds, eliminating unpaired electrons, the DOS plots reveal
nonmagnetic insulators {[}\figref{Bilayer_AB2}(c,e){]}.

\subsection{Step Edges in \emph{h}-BN\label{subsec:Steps}}

In real \emph{h}-BN sheets, it is difficult to manufacture bilayer
edges that are aligned to form a collection of stacked open N-edges
as in \figref{Bilayer_AB1}(b) although a preference for such alignment
has been observed in our study where edges were fabricated through
a high-energy electron beam \citep{dogan2020electron}. Thus, it is
important to investigate the behavior of a monolayer edge when its
neighbor does not also have an edge nearby. To this end, we have investigated
the behavior of step edges in \emph{h}-BN for AA, $\text{AA}^{\prime}$
and AB stacking sequences. We have found that the monolayer edges
remain intact and can be in an FM or AFM state. In \tabref{Steps},
we summarize these findings. We observe that the energy difference
between the FM and AFM states depends on the stacking sequence, and
hence the interaction energies in \eqref{Hamiltonian} for each edge
would be different. However, the FM state remains the ground state
and the AFM state is higher in energy by similar amounts, thus the
resulting models are not expected to be qualitatively different. Therefore,
the N-edge in a single layer should remain magnetically active when
it is sandwiched by other sheets that do not have an edge in the immediate
vicinity, regardless of the stacking sequence. 

\begin{table}
\def\arraystretch{2.0}
\begin{centering}
\begin{tabular}{c|c|c|c|c|}
\multicolumn{2}{c|}{} & $\ \ $NM$\ \ $ & $\ \ $FM$\ \ $ & $\ $AFM$\uparrow\downarrow$$\ $\tabularnewline
\hline 
\multirow{2}{*}{AA} & $\ $$\Delta E$ (meV)$\ $ & $\equiv0$ & $-164$ & $-85$\tabularnewline
\cline{2-5} 
 & $E_{gap}$(eV) & metal & $0.15$ & $0.31$\tabularnewline
\hline 
\multirow{2}{*}{$\text{AA}^{\prime}$} & $\Delta E$ & $\equiv0$ & $-197$ & $-152$\tabularnewline
\cline{2-5} 
 & $E_{gap}$ & metal & $0.12$ & $0.43$\tabularnewline
\hline 
\multirow{2}{*}{$\ $AB1$\ $} & $\Delta E$ & $\equiv0$ & $-180$ & $-115$\tabularnewline
\cline{2-5} 
 & $E_{gap}$ & metal & $0.18$ & $0.33$\tabularnewline
\hline 
\multirow{2}{*}{AB2} & $\Delta E$ & $\equiv0$ & $-206$ & $-163$\tabularnewline
\cline{2-5} 
 & $E_{gap}$ & metal & $0.13$ & $0.47$\tabularnewline
\hline 
\end{tabular}
\par\end{centering}
\caption{\label{tab:Steps}Total energies and electronic band gaps of the nonmagnetic
(NM), ferromagnetic (FM) and antiferromagnetic (AFM) configurations
of the N-terminated step edge of \emph{h}-BN in AA, $\text{AA}^{\prime}$
and AB stacking sequences. Total energies are given per edge N atom.}
\end{table}

\begin{figure*}
\centering{}\includegraphics[width=0.9\textwidth]{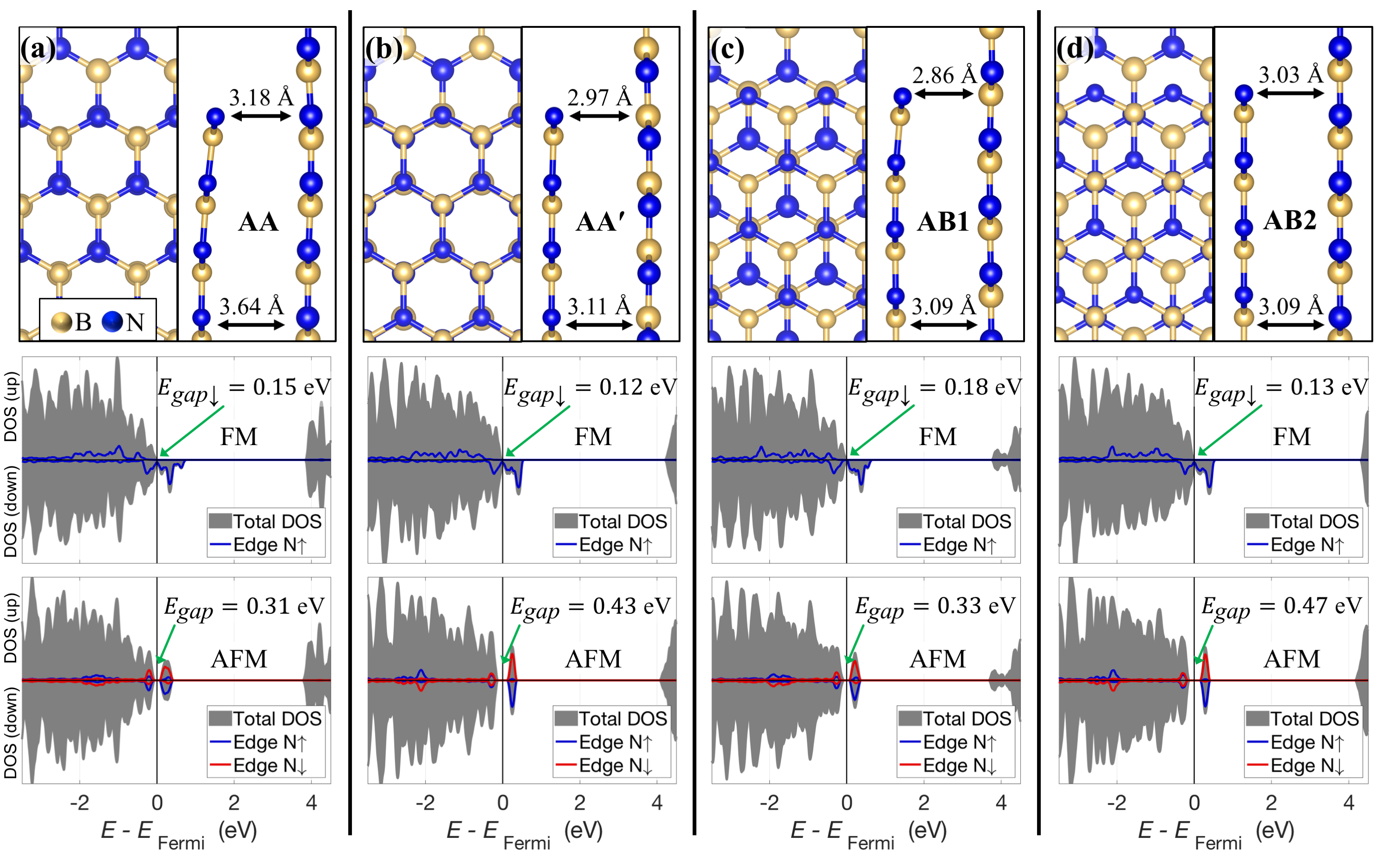}\caption{\label{fig:Bilayer_steps} Atomic and electronic structures of the
step edges in the AA (a), $\text{AA}^{\prime}$ (b), AB1 (c) and AB2
(d) stacking sequences. For each case, the spin-resolved densities-of-states
plots including the projections onto the orbitals of the edge nitrogen
atoms are presented for both the FM and AFM states. The values of
the electronic gaps, where they exist, are also printed on each panel.}
\end{figure*}

In \figref{Bilayer_steps}, we present the atomic and electronic structures
of the four step edges we have investigated. We find that the ground
state electronic structure of all four step edges is a FM half-semiconductor
with band gaps in the infrared region ($0.12-0.18$ eV), and the AFM
state is a nonmagnetic semiconductor with band gaps also in the infrared
region ($0.31-0.47$ eV). Our findings indicate that step edges can
be utilized to stabilize a magnetic edge in $\text{AA}^{\prime}$-\emph{h}-BN.
We also find that the amount of bending of the step edge toward the
full sheet is significantly different between the AB1 and AB2 stacking
sequences (0.23 Å and 0.06 Å, respectively), due to the alignment
of the edge N atom in the top layer with a B atom \emph{vs.} a hollow
site in the bottom layer. 

\section{Conclusion\label{sec:Conclusion4}}

We conducted a first-principles study of the commonly observed nitrogen-terminated
zigzag edges (N-edges) in \emph{h}-BN, with a focus on Bernal-stacked
\emph{h}-BN (AB-\emph{h}-BN). We showed that the N-edges in the monolayer
\emph{h}-BN possess various spin configurations that are within a
small total energy window. Because the magnetization is localized
in the vicinity of each N atom, we constructed a discrete lattice
model for magnetism on the edge, which yields nm-scale correlation
lengths at temperatures in the $10-50$ K range. Although in the long
edge limit ($N\rightarrow\infty$) no significant magnetization is
expected in external magnetic fields less than 1 Tesla, the shorter
edges that lie in the vertices of nanopores may exhibit significant
magnetization that may be probed by EELS or STM experiments. We showed
that these edges remain open in AB-\emph{h}-BN and can be stacked
in the out-of-plane direction to form bilayer or multilayer edges,
where the spins in neighboring layers do not interact significantly.
If, as we proposed, the spins in multilayer AB-\emph{h}-BN can be
manipulated \emph{via} magnetic fields, magnetoresistive effects as
well as spintronic applications may be achieved in these reduced-dimensional
systems. We also discussed the possibilities of closed edge reconstructions
in AB-\emph{h}-BN, as well as in $\text{AA}^{\prime}$-\emph{h}-BN
where they inevitably occur. Finally, we examined step edges in \emph{h}-BN
with various stacking sequences, which indicates that the edges in
a single layer are stable in all stacking sequences when they are
capped by full sheets. We hope that our results motivate experimental
studies that closely investigate the magnetic and electronic properties
of these edges in \emph{h}-BN.

\section*{Acknowledgements}

This work was supported by the Director, Office of Science, Office
of Basic Energy Sciences, Materials Sciences and Engineering Division,
of the U.S. Department of Energy under contract No. DE-AC02-05-CH11231,
within the sp2-bonded Materials Program (KC2207), which supported
first-principles computations of the atomic structures. Further support
for theoretical work was provided by the NSF Grant No. DMR-1926004
which supported first-principles computations of the precise electronic
structures. Computational resources were provided by the DOE at Lawrence
Berkeley National Laboratory\textquoteright s NERSC facility and the
NSF through XSEDE resources at NICS.

\bibliographystyle{apsrev4-1}
\bibliography{BN_edges}

\end{document}